\documentclass[12pt]{iopart}

\usepackage{iopams}  
\usepackage[T1]{fontenc}
\usepackage[latin9]{inputenc}
\usepackage{amssymb}
\usepackage{graphicx}
\usepackage{esint}
\usepackage{epsfig}

\begin{document}

\title{The effect of environmental coupling on tunneling of quasiparticles in Josephson junctions}

\author{Mohammad H. Ansari}

\address{Institute for Quantum Computing and Department of Physics and Astronomy,
University of Waterloo, 200 University Avenue West, Waterloo, ON,
N2L 3G1, Canada}

\author{Frank K. Wilhelm}

\ead{fwm@lusi.uni-sb.de}

\address{Institute for Quantum Computing and Department of Physics and Astronomy,
University of Waterloo, 200 University Avenue West, Waterloo, ON,
N2L 3G1, Canada}

\address{Theoretical Physics, Saarland University, 66123 Saarbrücken, Germany}

\author{Urbasi Sinha}

\address{Institute for Quantum Computing and Department of Physics and Astronomy,
University of Waterloo, 200 University Avenue West, Waterloo, ON,
N2L 3G1, Canada}

\address{Raman Research Institute, Sadashivanagar, Bangalore 560080, India.}

\author{Aninda Sinha}

\address{Centre for High Energy Physics, Indian Institute of Science, Bangalore,
India}

\begin{abstract}
We study quasiparticle tunneling in Josephson tunnel junctions embedded
in an electromagnetic environment. We identify tunneling processes
that transfer electrical charge and couple to the environment in a
way similar to that of normal electrons, and processes that mix electrons
and holes and are thus creating charge superpositions. The latter
are sensitive to the phase difference between the superconductor and
are thus limited by phase diffusion even at zero temperature. We show that the environmental coupling is suppressed in many environments, thus leading to
lower quasiparticle decay rates and thus better superconductor qubit
coherence than previously expected. Our approach is nonperturbative in
the environmental coupling strength.

\end{abstract}


\maketitle

\section{Introduction}

The physics of micro and nanoscale Josephson junctions is a paradigmatic
application of macroscopic dissipative quantum mechanics of open systems
\cite{{Caldeira81},{Caldeira83},{Ingold92}}. This, on the one hand,
makes them ideal test-beds for that theory, with largely tunable
parameters \cite{{Kycia01},{PRL01}}. On the other hand, Josephson
junctions have an ample range of applications in sensing \cite{Clarke04},
amplification \cite{Mueck03}, metrology \cite{Jeanneret01} and quantum
information processing \cite{{Insight},{You05b},{Shumeiko06},{Makhlin01}}.
For these applications, it is imperative to thoroughly understand
the dissipative quantum physics of Josephson Junctions in order to
achieve optimal device performance.

Over a long time, work on this topic has focused on the {\em transport}
properties of Josephson junctions, their current-voltage and noise
characteristics \cite{Ansari:2011gj}. The `P(E)'-theory of treating environmental fluctuations
\cite{{Ingold92},{Schoen98}} is established as a powerful tool in the
case of small junctions when (Josephson or quasiparticle) tunneling
can be treated perturbatively.

For applications in quantum computing, another question in this framework
has emerged \cite{{Martinis09},{Martinis09b}}: Rather than computing
current-voltage characteri
stic, the total quasiparticle transition
rate is the quantity of interest. This is important because quasiparticle
transitions in any direction are highly detrimental to qubit coherence
\cite{Lutchyn06}. References {[}\cite{{Lutchyn06},{Martinis09},{Martinis09b},{Catelani11},{Leppakangas11}}{]}
study this problem in an approach that is perturbative in the coupling
to the environment as measured by the environmental impedance in units
of the quantum resistance $R_{Q}=h^{2}/e\simeq25.8\; k\Omega$. In
this paper we are going to generalize that study to arbitrary system-environment
interaction strength. Some of the processes
mix the electron-like and hole-like branches of the quasiparticle
spectrum and depend on the phase difference between the
superconductors. These are generally phase-dependent but highly
sensitive to the environment. These observations lead to slower relaxation than
expected in previous work.

The paper is organized as follows: We introduce the mathematical
formulation of the tunneling model in section \ref{sec:model}. We will
compute the tunnling rate in lowest order of the tunnel coupling but
to all orders in the coupling of the quasiparticles to the environment
in section \ref{sec:rate}, which can be a
general linear impedance.  We specialize on it being an undamped
harmonic oscillator  in section \ref{sec:LCimpedance}, which directly relates to
Refs. [\cite{{Martinis09},{Martinis09b},{Catelani11},{Leppakangas11}}]
but shows that the phase-dependent component is usually reduced due to
dressing by zero-point fluctuations. We discuss and overdamped
environment that may occur in other applications in section \ref{sec:overdamped}.

\section{Mathematical formulation}

\begin{figure}
\includegraphics[scale=0.23]{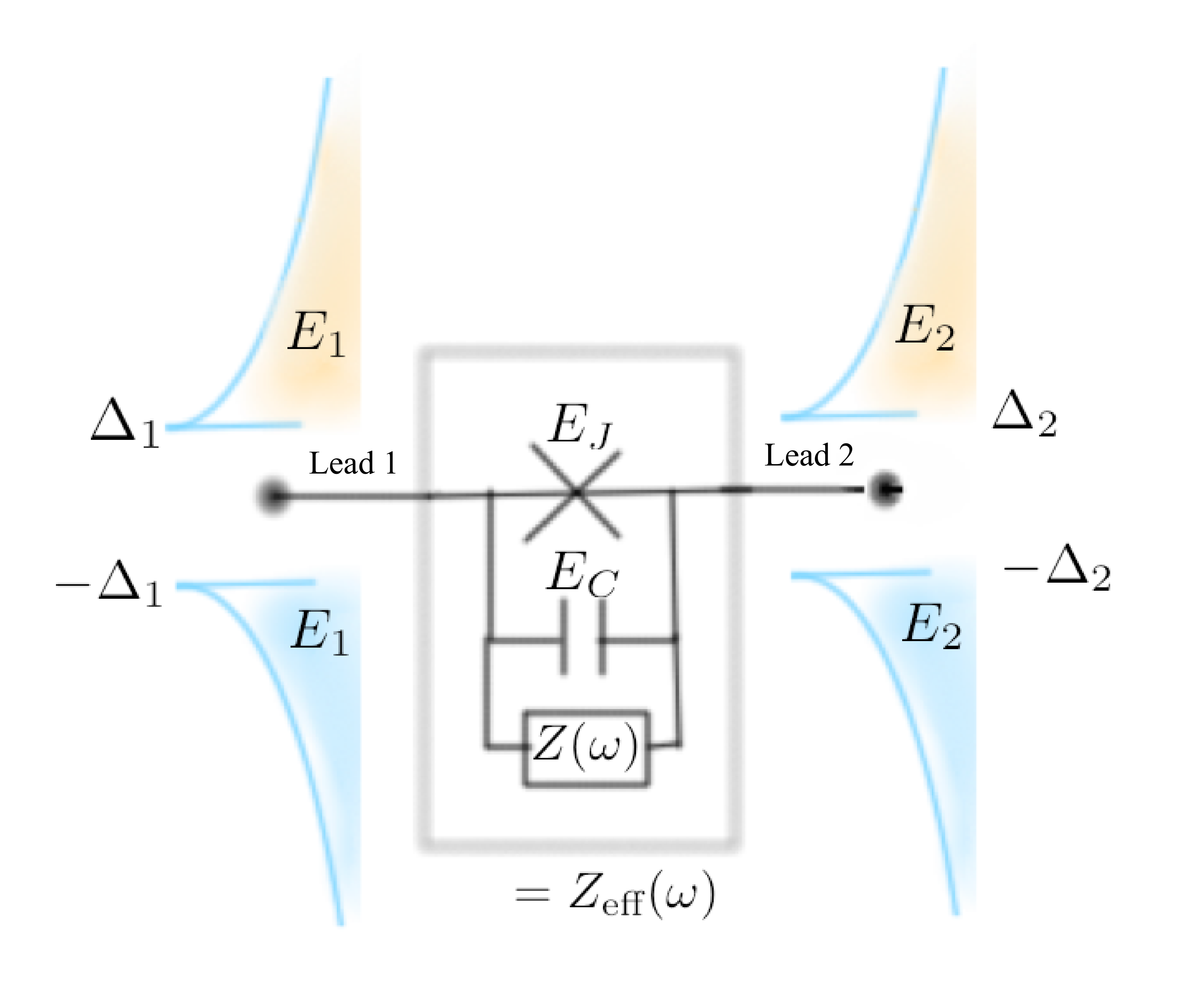} 
\caption{Quasiparticle processes labeled by their dispersion connected to an external circuit. By linearization, the Josephson circuit is replaced
by an effective impedance.
\label{fig:circuit}}
\end{figure}

\subsection{Model \label{sec:model}}

We start from the Hamiltonian 
\begin{equation}
\hat{H}=\hat{H}_{{\rm BCS,1}}+\hat{H}_{{\rm BCS,2}}+\hat{H}_{T}+\hat{H}_{{\rm env}}+\hat{H}_{{\rm env-c}}.
\end{equation}

Here, we have the BCS Hamiltonians \cite{{deGennes66},{Bardeen57}}in
mean field form describing the two electrodes

\begin{eqnarray}
\hat{H}_{BCS,i} & = & \sum_{k,\sigma}\left(\xi_{k}-\mu_{i}\right)\hat{c}_{k,\sigma,i}^{\dagger}\hat{c}_{k,\sigma,i}\label{eq:BCS_Ham}\\
 &  & +\Delta_{i}\sum_{k}\hat{c}_{k,\uparrow,i}\hat{c}_{k,\downarrow,i}+\Delta_{i}^{\ast}\sum_{i}\hat{c}_{k,\uparrow,i}^{\dagger}\hat{c}_{k,\downarrow,i}^{\dagger}\nonumber 
\end{eqnarray}
where $\xi_{k}=\frac{\hbar^{2}k^{2}}{2m}$ is the kinetic energy of
the electron, $\mu_{i}$ is the chemical potential in superconductor
$i=1,2$, being $\mu_{1}=E_{F}$ and $\mu_{2}=E_{F}+eV$ where $V$
is the applied voltage. We keep $\Delta_{i}=\left|\Delta\right|e^{i\phi_{i}/2}$
complex in order to allow for a phase difference. The tunneling Hamiltonian
is 
\begin{equation}
\hat{H}_{T}=\sum_{kl}\left(T_{kl}\hat{c}_{k,\sigma,1}^{\dagger}\hat{c}_{l,\sigma,2}+T_{kl}^{\ast}\hat{c}_{k,\sigma,2}^{\dagger}\hat{c}_{l,\sigma,1}\right)\label{eq:tunneling}
\end{equation}

Before discussing the electromagnetic environment, we diagonalize
the BCS Hamiltonians, eq. (\ref{eq:BCS_Ham}) through the Bogoliubov
transformation: 
\begin{eqnarray*}
\hat{c}_{k\uparrow} & = & u_{k}^{*}\hat{\gamma}_{k\uparrow}+v_{k}\hat{\gamma}_{-k\downarrow}^{\dagger},\\
\hat{c}_{-k\downarrow}^{\dagger} & = & -v_{k}^{*}\hat{\gamma}_{k\uparrow}+u_{k}\hat{\gamma}_{-k\downarrow}^{\dagger}.
\end{eqnarray*}

Here, the BCS coherence factors are defined 
\begin{eqnarray}
u_{k,i} & =\sqrt{\frac{1}{2}\left(1+\frac{\xi_{k,i}-\mu_{i}}{E_{k,i}}\right)}e^{i\phi_{i}/2}\label{eq:BCS_u}\\
v_{k,i} & =\sqrt{\frac{1}{2}\left(1-\frac{\xi_{k,i}-\mu_{i}}{E_{k,i}}\right)}e^{-i\phi_{i}/2}\label{eq:BCS_v}
\end{eqnarray}
where we introduced the quasiparticle energy $E_{k,i}=\sqrt{(\xi_{k,i}-\mu_{i})^{2}+|\Delta|^{2}}$.
This allows us to rewrite the tunneling Hamiltonian, eq. \ref{eq:tunneling}
in the following form

\begin{eqnarray}
\hat{H}_{T} & = & \sum_{k,\sigma=\uparrow,\downarrow}\left(T_{kl}|u_{k1}u_{l2}|e^{i\phi/2}-T_{-l-k}^{*}|v_{k1}v_{l2}|e^{-i\phi/2}\right)\nonumber \\
 &  & \times\left(\hat{\gamma}_{k\sigma1}^{\dagger}\hat{\gamma}_{k\sigma2}+\hat{\gamma}_{-k\sigma2}^{\dagger}\hat{\gamma}_{-l\sigma1}\right)+H_{T2}\label{eq. HT}
\end{eqnarray}
where we introduced the phase difference $\phi=\phi_{1}-\phi_{2}$.
The term $\hat{H}_{T2}$ contains operators that change the number
of quasiparticles by two, i.e., contain terms of the structure $\hat{\gamma}\hat{\gamma}$
and $\hat{\gamma}^{\dagger}\hat{\gamma}^{\dagger}$ hence changing
the total number of quasiparticles in the setup, which do not contribute
to the quasiparticle rate --these terms contribute to Josephson and
Andreev processes. We now want to evaluate the Fermi's golden rule
rate for a transition that transfers a quasiparticle from the electrode
1 to the electrode 2. The relevant matrix element is 
\begin{eqnarray*}
\langle\left(N_{1}-1\right)_{k\sigma},\left(N_{2}+1\right)_{l\sigma}\left|\hat{H}_{T}\right|N_{1k\sigma},N_{2l\sigma}\rangle=\\
T_{kl}e^{i\phi/2}\left|u_{k1}u_{l2}\right|-T_{-k-l}^{*}e^{-i\phi/2}\left|v_{k1}v_{l2}\right|
\end{eqnarray*}

\begin{figure}[h]
\includegraphics[width=0.4\columnwidth]{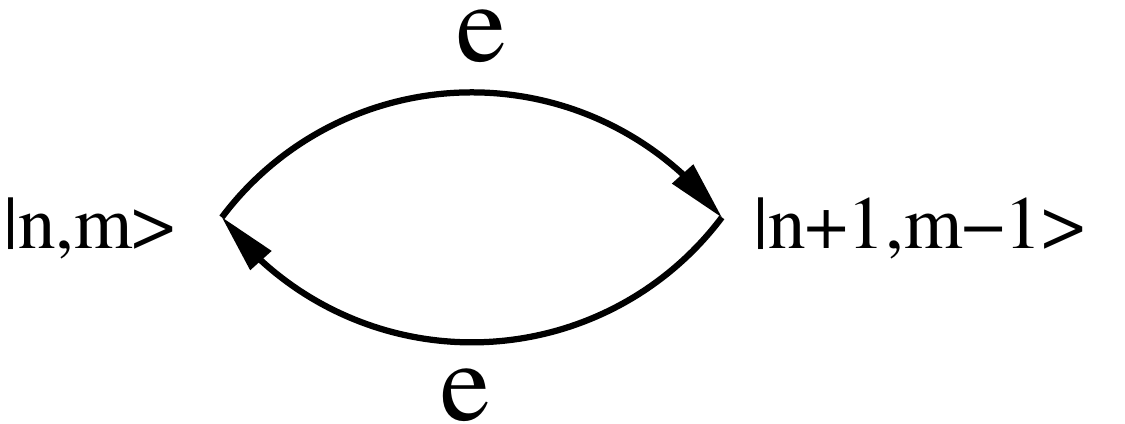}\quad{}\includegraphics[width=0.5\columnwidth]{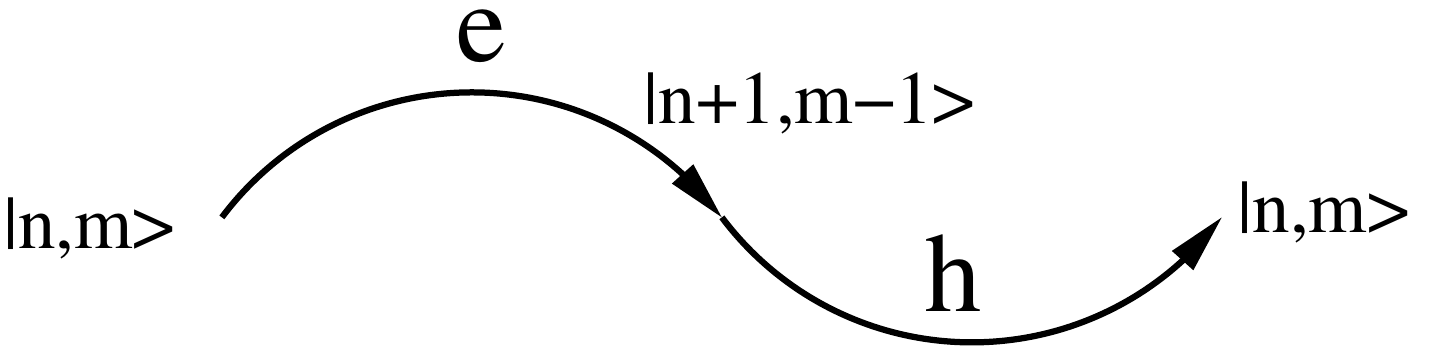}
\caption{\label{fig:chargediagrams} Diagrammatic representation of the different
contributions to the quasiparticle tunneling matrix elements. Left:
Processes closed in quasiparticle and in charge space, right: quasiparticle
process that creates charge superpositions. The typical state $|n,m\rangle$
represents $n$ particles in the lead 1 and $m$ particles in the
lead 2.}
\end{figure}

We now use that for a nonmagnetic barrier, $T_{kl}=T_{-k-l}^{*}$
can be chosen real. In the Fermi golden rule transition rate, we
need the absolute square 
\begin{eqnarray*}
\left|\langle\left(N_{1}-1\right)_{k\sigma},\left(N_{2}+1\right)_{l\sigma}\left|\hat{H}_{T}\right|N_{1k\sigma},N_{2l\sigma}\rangle\right|^{2}\\
=T_{kl}^{2}\left(|u|^{2}+|v|^{2}-2|uv|\cos\phi\right)
\end{eqnarray*}
where $u(E,E^{\prime})=|u_{1}(E)u_{2}(E^{\prime})|$ and $v=|v_{1}(E)v_{2}(E^{\prime})|$
with $u_{1/2}(E)$ and $v_{1/2}(E)$ given by the BCS coherence formulae,
eqs. (\ref{eq:BCS_u}) and (\ref{eq:BCS_v}) Thus, the transition
probability contains an interference term that is sensitive to the
phase across the junction. This term describes a process that, even
though it transfers a genuine quasiparticle from electrode 1 to electrode
2, it actually consists of a superposition of electron and hole transfer,
i.e., it is not diagonal in charge space, see the diagrammatic representation
in fig. \ref{fig:chargediagrams}. We will see later on how this is
important in the sensitivity to environment. This term bears analogy
to the famous $\cos\phi$ term in the Josephson effect \cite{{Barone82},{Catelani:2012we},{Catelani:2011cf},{catelani_2011_QpRelax}}.

\subsection{Tunneling rate \label{sec:rate}}

In order to capture the influence of the environment, we apply the
ideas of $P(E)$-Theory \cite{Ingold92}. There, the environmental
Hamiltonian is described by an oscillator bath $\hat{H}_{{\rm bath}}=\sum_{n}\omega_{n}\hat{a}_{n}^{\dagger}\hat{a_{n}}$
and couple the oscillators linearly to our quasiparticle system $\hat{H}_{env-c}=\hat{N}_{1}e\delta\hat{V}$,
where we assume that only the first reservoir fluctuates - this corresponds
to a specific choice of gauge. The total number and voltage operators
are 
\begin{equation}
\hat{N}_{1}=\sum_{k,\sigma}\hat{c}_{k\sigma1}^{\dagger}\hat{c}_{k\sigma1},\quad\delta\hat{V}=\sum\lambda_{i}\left(\hat{a}_{i}+\hat{a}_{i}^{\dagger}\right).
\end{equation}
We can now work out the total tunneling rate by summing over momentum
states at constant energy from the initial state $|i\rangle=|R\rangle|E\rangle$
to the final state $|f\rangle=|R'\rangle|E'\rangle$ and considering
that the matrix element of environmental Hamiltonian $\langle E|T_{kq}\gamma_{q\sigma}^{1\dagger}\gamma_{k\sigma}^{2}|E'\rangle$
 thermal distribution of quasiparticles. The transition
rate formally can be written

\begin{eqnarray}
\vec{\Gamma}_{1}\left(V\right)=\frac{1}{e^2 R_N}\int_{-\infty}^{\infty}dEdE^{\prime}\, D_{1}(E)D_{2}(E^{\prime})\nonumber \\
~~~~~~ \times f_{1}(E)[1-f_{2}(E^{\prime})]P_{{\rm tot}}(E,E^{\prime})\label{eq: rate with Ptot}
\end{eqnarray}
where $D_{i}$ is the reduced density of states  and $f_{i}$ is the
corresponding distribution functions on the lead $i$. We consider
here that the nonequilibrium quasiparticle tunneling keeps the system
in thermal and chemical-potential equilibrium. 
The influence
of the external circuitry on quasiparticle tunneling is encoded in
the dressed vertex $P_{{\rm tot}}(\epsilon)$, the probability density
for exchanging energy $\epsilon$ with the environment. From the Fermi's
golden rule the tunneling transition rate can be written in more detail
as 

\begin{eqnarray*}
&& \vec{\Gamma}_{1}(V)  =  \frac{1}{e^{2}R_{N}}\int dEdE'D_{1}(E)D_{2}(E')f_{1}(E)(1-f_{2}(E')) \\
&& \times \sum_{R,R'}\left|\left\langle R'|H_{T}|R\right\rangle \right|{}^{2}P_{\beta}(R)\delta(E+eV-E'+E_{R}-E{}_{R'}),
\end{eqnarray*}
where $P_{\beta}(R)=\langle R|\rho_{\beta}|R\rangle$ for
$\rho_{\beta}=e^{-\beta H_{\rm env}}/\sum e^{-\beta H_{\rm env}}$ and the
arrow on $\Gamma$ indicates the direction of tunneling from the lead
1 to 2. $R_{N}$ is the tunnel resistance of the junction in the normal
state. \footnote{By expanding the Dirac delta as a temporal integral $\delta(x)=\left(2\pi\hbar\right)^{-1}\int_{-\infty}^{\infty}dte^{-\frac{i}{\hbar}xt}$
and writing the operator $\langle R'|e^{-E_{R'}t}\hat{H}_{T}e^{E_{R}t}|R\rangle=\langle R'|\hat{H}_{T}(t)|R\rangle$
the time evolution of the Hamiltonian makes the rate to become proportional
to $\langle\langle H_{T}\left(t\right)H_{T}\left(0\right)\rangle\rangle$,
where $\langle\langle\cdots\rangle\rangle=\sum_{R,R'}\left\langle R'\left|\cdots\rho_{\beta}\right|R\right\rangle $.
Substituting the tunneling Hamiltonian into this formula one gets
$\left\langle\left\langle H_{T}\left(t\right)H_{T}\left(0\right)\right\rangle\right\rangle=\left\langle\left\langle\left(ue^{i\phi(t)/2}-ve^{-i\phi(t)/2}\right)\left(ue^{-i\phi(0)/2}-ve^{i\phi(0)/2}\right)\right\rangle\right\rangle$.
Reordering using properties of Gaussian states leads to the following
expressions  
$\langle e^{\pm i\phi(t)}e^{\mp i\phi(0)}\rangle\sim e^{\langle(\phi(t)-\phi(0))\phi(0)\rangle}$
and $\langle e^{\pm i\phi(t)}e^{\pm i\phi(0)}\rangle\sim e^{-\langle(\phi(t)+\phi(0))\phi(0)\rangle}.$}   Considering that the phase has fluctuations around the classical value $\phi=\varphi+\delta \phi$ and substituting these all  one can get the transition rate

\begin{eqnarray}
\vec{\Gamma_{1}}(V) & = & \frac{4}{e^{2}R_{N}}\int dEdE'\int\frac{dt}{2\pi\hbar}e^{\frac{i}{\hbar}\left(E+eV-E'\right)}D_{1}(E)D_{2}(E^{\prime})f_{1}(E)(1-f_{2}(E'))\nonumber \\
 &  & \times\left(\left(u^{2}+v^{2}\right)e^{\frac{\langle\delta \phi(t)\delta \phi(0)\rangle}{4}}-2uv\cos \varphi e^{-\frac{\langle\delta \phi(t) \delta \phi(0)\rangle}{4}}\right)e^{-\frac{\langle \delta \phi(0) \delta \phi(0)\rangle}{4}}\label{eq: rate final u and v}
\end{eqnarray}
where $u(E,E^{\prime})=|u_{1}(E)u_{2}(E^{\prime})|$
and $v=|v_{1}(E)v_{2}(E^{\prime})|$ with $u_{1/2}(E)$ and $v_{1/2}(E)$
given by the BCS formulae, eqs. (\ref{eq:BCS_u}) and (\ref{eq:BCS_v}).

Note the the phase fluctuations defined here by the effective voltage across the junction 
\begin{equation}
\frac{\delta\phi(t)}{2}=\frac{e}{\hbar}\int_{0}^{t}dt^{\prime}\,\delta V(t^{\prime}). \label{eq. def. phi}
\end{equation}
is the conjugate to the Cooper pair charge. The convention used here is consistent with
  assuming the resistance quantum to be $R_K/4=h/4e^2$.

{ The vertex probability is defined as the probability density for exchanging
energy $\epsilon$ between system and the environment and is defined
formally as 

\begin{eqnarray}
P_{{\rm tot}}(E,E^{\prime}) & = \int_{-\infty}^{\infty}\frac{dt}{2\pi\hbar}\, e^{J(t)+\frac{i}{\hbar}(E+eV-E^{\prime})t}\label{eq:ptot}
\end{eqnarray}
By comparing eq. (\ref{eq: rate with Ptot}) and (\ref{eq: rate final u and v})
one can correctly expect the definition $e^{J(t)}=\left(u^{2}+v^{2}\right)e^{J_{s}(t)}-2uv\cos \varphi e^{J_{a}(t)}$,
where $e^{J_{s}(t)}=\langle e^{i\delta \phi(t)/2}e^{-i \delta \phi(0)/2}\rangle=e^{\langle\left(\delta \phi(t)-\delta \phi(0)\right)\delta \phi(0)\rangle/4}$
and $e^{J_{a}(t)}=\langle e^{i\delta \phi(t)/2}e^{i\delta \phi(0)/2}\rangle= e^{-\langle\left(\delta \phi(t)+\delta \phi(0)\right)\delta \phi(0)\rangle/4}$.
From these equalities one can simplify the definitions to $J_{s}(t)=\left\langle (\delta\phi(t)-\delta\phi(0))\delta\phi(0)\right\rangle /4$
and $J_{a}(t)=-\left\langle
  (\delta\phi(t)+\delta\phi(0))\delta\phi(0)\right\rangle/4$. In this
notation, one needs to pay attention to the observation that the to
whereas $J_{s/a}(t)$ are purely properties of the environment, the
combined quantity $\exp(-J(t))$ inevitably contains coherence factors of
the superconductor. Concurrently, we decompose $P_{\rm tot}$ into two contributions
\begin{equation}
P_{s/a}(E)=\int \frac{dt}{2 \pi \hbar}\, e^{iEt}e^{J_{s/a}(t)}.\label{eq: Ps}.
\end{equation}
Note that in the definitions we considered that the mean value of
the phase is set by the external bias and the effect of environment
is summarized into the presence of phase fluctuations about this
mean value. The symmetric combination $J_{s}(t)$ captures all-electron
and all-hole processes and the corresponding vertex $P_s(E)$ coincides with
the all-electron $P(E)$ known from Ref.{[}\cite{Ingold92}{]}

However since  $J_{a}(t)=\left\langle (\delta\phi(t)+\delta\phi(0))\delta\phi(0)\right\rangle /4=J_{s}(t)+\left\langle \delta\phi^{2}(0)\right\rangle /2$
captures processes that mix electrons and holes.  Although $J_{s}(0)=0$ the asymmetric correlation function does not vanish at
$t=0$ as it becomes $J_{a}(0)=\left\langle \delta\phi^{2}(0)\right\rangle /2$
captures zero-point fluctuation. $P_{s}(E)$ is a normalized probability as $\int  P_s(E)
dE=\exp(-J_s(0))=1$; however, $P_{a}(E)$ is not normalized
$\int P_a(E)dE = \exp(-J_a(0))=\exp \langle \delta \phi^2 (0)/2
\rangle $ as it describes electron-hole coherence for which no
conservation law should be expected.  We can understand this as follows: Even
though electron-hole mixing processes are diagonal in the Fock space
of quasiparticles, they are off-diagonal in charge space. The electromagnetic
noise couples to these electrical charges. The environmental modes
dress the charge in an attempt to measure charge and localize it.
This has an  analogy: Creation of a superposition of charge
states occurs, following the charge-flux uncertainty relation, whenever
the phase across a Josephson junction is becoming localized, i.e.,
when a Josephson junction is behaving classically. Thus, on the one
hand, the scaling factor $\exp \langle \delta \phi^2 (0)/2
\rangle $, directly measures
the degree of charge localization. On the other hand, research aiming
at maximum supercurrent (hence maximally localized phase and maximally
extended charge) in small Josephson junctions in a dissipative environment
finds $\exp(-\langle\phi(0)^{2}\rangle/2)$ to be the relevant reduction
factor \cite{{Steinbach01},{Joyez99}}.

In order to compute $J_{s/a}$  introduce an oscillator
bath model and match its properties to the fluctuation-dissipation
theorem as it describes Johnson-Nyquist noise. Applying the standard
procedures used in $P(E)$-theory, we find 

\begin{equation}
S(t)=\left\langle \delta\phi(t)\delta\phi(0) \right\rangle =2\int_{-\infty}^{\infty}\frac{d\omega}{\omega}\frac{{\rm Re}Z_{{\rm eff}}(\omega)}{R_{K}}\frac{e^{-i\omega t}}{1-e^{-\beta\hbar\omega}}
\end{equation}

{ Notably, this integral seems to diverge at its infrared end, where
it formally appears to be $\simeq\int\frac{d\omega}{\omega}Z_{{\rm eff}}(0)$
at $T=0$ and $\propto T\int\frac{d\omega}{\omega^{2}}{\rm Re}Z_{{\rm eff}}(0)$
at $T>0$. In the combination $4J_{s}(t)=S(t)-S(0)$ this divergence
is removed, but the same argument does not apply to $4J_{a}(t)=S(t)+S(0)$.
Thus, $Z_{{\rm eff}}(0)\neq0$, enforces
$P_a=0$, meaning that all electron-hole mixed processes would disappear.
We will see later that this does not occur in standard physical environments. }

The final expression for the probability density $P(E,E^{\prime})$
is thus

\begin{eqnarray}
P_{{\rm tot}}(E,E^{\prime}) & = & \int\frac{dt}{2\pi\hbar}\, e^{-i(E+eV-E')t}e^{-\frac{S(0)}{4}}\label{eq: ptot_final}\\
 &  & \times\left[(u^{2}+v^{2})e^\frac{S(t)}{4}-2uv\cos \varphi
   e^{-\frac{S(t)}{4}} \right]
\nonumber
\end{eqnarray}

\section{Models for the environment}

In order to evaluate the typical $P_{s/a}(E)$ and the resulting tunneling
rates, we need to project likely models of the electromagnetic environment
providing the impedance $Z_{{\rm eff}}$. Following the resistively and capacitatively shunted junction (RCSJ) models and its microscopic
analogues \cite{Tinkham96,Ambegaokar82,Caldeira83,Caldeira81}, we
consider the quasiparticle channel to be put in parallel to the supercurrent,
Josephson channel, the junction capacitance, and an external impedance
$Z(\omega)$, see Fig. (\ref{fig:circuit}). We model the linear impedance
of the Josephson channel by its Josephson inductance $L_{J}=\Phi_{0}/(2\pi I_{c})$.
The total effective impedance of this parallel setup leads to 
\begin{equation}
{\rm Re}Z_{{\rm eff}}=\frac{\omega^{2}{\rm Re}\left(Z\right)}{C^{2}|Z|^{2}}\frac{1}{\left(\omega^{2}-\omega_{p}^{2}\right)^{2}+\omega\omega_{P}^{3}\frac{{\rm Im}Z}{|Z|^{2}}+\frac{\omega^{2}}{C^{2}|Z|^{2}}}
\end{equation}
with $\omega_{p}=(CL_{J})^{-1/2}$ is the junction's plasma frequency.
In the case of a simple resistor or lossless transmission line, $Z(\omega)=R$
and we find 
\begin{equation}
{\rm Re}[Z_{{\rm eff}}]=\frac{\omega^{2}}{RC^{2}}\frac{1}{(\omega^{2}-\omega_{p}^{2})^{2}+\omega^{2}/(RC)^{2}}\label{eq:lorentz_spectrum}
\end{equation}
Here, we can observe that at low frequencies, ${\rm Re}Z_{{\rm eff}}\propto\omega^{2}$.
Physically, the reason for this is that the noise from the resistor
gets shunted through the Josephson junction at low frequencies and
does not affect the quasiparticle channel. Thus, although $P_{a}$
will be normalized to a value smaller than unity, the scenario of
it scaling all the way to zero does not occur. This is consistent
with the fact that phase qubits do show a zero-voltage current. The
same general conclusion holds for other physical environments tested.

\subsection{Infinite-quality environmental mode \label{sec:LCimpedance}}

Given the high quality of qubit junctions, it is appropriate to start
from the limit of an infinite quality factor, $R\rightarrow\infty$.
In that case, the effective impedance reduces to 
\begin{equation}
Z_{{\rm eff}}=\frac{\pi}{2C}\left[\delta\left(\omega-\omega_{p}\right)+\delta\left(\omega+\omega_{p}\right)\right].
\label{eq. Z}
\end{equation}

 We introduce the dimensionless parameter wave impedance 
$\rho_c=4\pi Z_{0}/R_{K}$,
where $Z_{0}=\sqrt{L_{J}/C}$, which emerges from eq. (\ref{eq. Z})
and $\omega_{p}=1/\sqrt{L_{J}C}$ and $L_{J}=\Phi_{0}/2\pi I_{c}$. The
index $c$ indicates that in this definition the cooper pair quantum of
resistance $R_K/4$ was considered, different from Ref.
{[}\cite{Ingold92}{]}.  The Cooper pair vacuum fluctuation is controlled by its corresponding wave impedance $S(0)=\rho_c$.  The total environmental
transition probability is defined 
\begin{equation}
P(E)=\sum_{k=-\infty}^{\infty} p_{k}(\rho_c,\omega_{P,}T)\delta(\omega-k\omega_{p}),
\end{equation}
which has the form of a series of sidebands, corresponding to the
emission / absorption of $k$ photons to/from the plasma mode.   The
weights in the low temperature limit of $k_{B}T\ll\hbar\omega_{p}$
is Poissonian $p_{k}\left(\rho,\omega_{p}\right)=e^{-\rho_c/4}(\rho_c/4)^{k}/k!.$
This coincides with that of Ref. {[}\cite{Martinis09b}{]} in the limit of $\rho_c\rightarrow0$.
The total quasiparticle tunneling rate hence reads

\begin{eqnarray*}
\vec{\Gamma}_{1} &=&   \sum_{n=-\infty}^{\infty}\Gamma_{1n} \\ &=& \frac{4}{R_{T}e^{2}}\sum_{n=-\infty}^{\infty}\int_{\Delta}^{\infty}dE\, D_{1}(E)D_{2}(E+n\hbar\omega)f_{1}(E)(1-f_{2}(E+n\hbar\omega))\\
 &  & \times\left[\left(u^{2}(E,E+n\hbar\omega)+v^{2}(E,E+n\hbar\omega)\right)e^{\frac{S(t)}{4}} \right. \\ && \left. -2u(E,E+n\hbar\omega)v(E,E+n\hbar\omega)\cos \varphi e^{-\frac{S(t)}{4}}\right]e^{-{\frac{\rho_c}{4}}}p_{n}(\rho,\omega,T)
\end{eqnarray*}

For large capacitance junctions one can Taylor expand in small phase
fluctuations  $\left(u^{2}+v^{2}\right)\exp\left[S(t)/4\right]-2uv\cos \varphi \exp\left[-S(t)/4\right]\sim\left(u-v\right)^{2}+\left(u+v\right)^{2}S(t)/4$.
Considering that the phase fluctuation is small and stable, i.e. $S(t)\approx S(0)=\rho_{c}$ the final result is consistent with the results taken
from Ref. {[}\cite{{Martinis09},{Catelani:2011cf}}{]}  except that ours show that there is a
prefactor $\exp(-\langle\delta\phi(0)^{2}\rangle/4)$ in the rate
originating from zero-point fluctuations
that can reduce the quasiparticle
environmentally mediated transition rate.

In order to finally compute the rate, we use eq. (\ref{eq:BCS_u})
and (\ref{eq:BCS_v}) one can show that 
\begin{equation}
\left(u_{1}u_{2}\right)^{2}+\left(v_{1}v_{2}\right)^{2}  =  \frac{1}{2}\left(1+\frac{\xi_{1}\xi_{2}}{E_{1}E_{2}}\right) , 
u_{1}u_{2}v_{1}v_{2}  =  \frac{\Delta_1 \Delta_2}{4 E_1E_2}
\label{eq: u2+v2 and uv (E)}
\end{equation}

Substituting the density of states $D(E)=E/\xi=E/\sqrt{ E^2-\Delta^2}$ and the coherence factors in the
transition rate, and ignoring the pure thermal transition we arrive at the complete formula for
nonequilibrium quasiparticle tunneling

\begin{eqnarray} 
\nonumber \vec{\Gamma}_{1}  =   \sum_{n=-\infty}^{\infty} e^{-\frac{\rho_c}{2}} \frac{(\frac{\rho_c}{4})^n}{n!} && \left( \frac{\Gamma_{\textrm{bare}}}{2}  + \frac{2}{e^{2}R_{N}}  \int_{\Delta}^{\infty}dE\, f_{1}(E)(1-f_{2}(E+n\hbar\omega))    \right. \\ & & \left. \frac{E\left(E+n\hbar\omega\right)e^{{S(t)}/4}-\Delta_{1}\Delta_{2}\cos \varphi e^{-{S(t)/4}}}{\sqrt{\left(E^{2}-\Delta_{1}^{2}\right)\left(\left(E+n\hbar\omega\right)^{2}-\Delta_{2}^{2}\right)}}  \right)\\
\label{eq: rate complete formula-1}
\end{eqnarray}

where $ \Gamma_{{\rm bare}}=(4/R_{N}e^{2}) \int_{\Delta}^{\infty}dE\,
f_{1}(E)(1-f_{2}(E))$ and corresponds to the normal electron tunneling
rate  under a bias voltage and the second term in eq. (\ref{eq: rate
  complete formula-1}) is dressed term rooted from the quasiparticle
tunneling.  We see several nonperturbative features in this
expression. One is, of course, the occurrence of higher-order sidebands
corresponding to exchange of multiple photons with the
environment. The other one is that, as a consequence of normalization
of the total probability, even the
single-photon peak obtains a nonpeturbative weight factor.  In the limit of large superconducting gap $\Delta \gg \hbar \omega $ and a large capacitance junction, one particle exchange at low-lying quasiparticle levels  indicates the  total one particle rate to become

\begin{eqnarray} \nonumber
\Gamma_1 =   e^{-\frac{\rho_c}{2}} \frac{\rho_c}{4} \Gamma_{\textrm{bare}}  + \frac{2}{e^{2}R_{N}} e^{-\frac{\rho_c}{2}} \left(\frac{\rho_c}{4}\right)^2 && \int_{\Delta}^{\infty}dE\,  f_{1}(E)(1-f_{2}(E+\hbar\omega))  \\  & & \frac{E\left(E+\hbar\omega\right)+\Delta_{1}\Delta_{2} }{\sqrt{\left(E^{2}-\Delta_{1}^{2}\right)\left(\left(E+\hbar\omega\right)^{2}-\Delta_{2}^{2}\right)}} ,
\label{eq. gamma1}
\end{eqnarray}
where the second term in the parenthesis of eq. (\ref{eq. gamma1}) is
in fact the dressed tunneling rate $\Gamma_{\rm dr,1}$. Thus, we
directly see the reduction of the phase-sensitive tunneling term by
zero-point fluctuations described above made quantitative.

This rate depends on the details of the energy distribution
function. One of us \cite{ansari13} derived the explicit temperature
dependence of this rate out of equilibrium. For a junction with small macroscopic phase  in thermal equilibrium at low temperatures $T \ll \Delta$,  by substituting the Fermi-Dirac distribution function, the rate is

\begin{equation}\nonumber
\Gamma_{1}=\frac{\Delta }{2e^{2}R_{N}}{\rho_{c}}  \exp \left(\frac{E_{if}}{2k_BT }-\frac{\Delta}{k_BT}-\frac{\rho_{c}}{2}\right) K_{0}\left(\frac{E_{if}}{2k_BT }\right)
\label{eq. appA final} 
\end{equation} 
where $E_{if}$ is the parity transition energy in the qubit from odd
to even states and $K_0$ is a Bessel function.  This result is different from that of
Ref. \cite{Catelani:2011cf} (see eq. 35) by the dressing factor
$\exp\left({-\sqrt{E_c/2E_J}}\right) $.  More general formulation for
arbitrary phase $\varphi$ are worked out in Ref. \cite{ansari13}.

For phase qubits,
whose impedance is engineered to be $Z_{0}\simeq50\Omega$ this correction
does not qualitatively change the physics, however, improves the quality
of fits to the data. In other type of qubits with higher impedance
these corrections will be crucial. A qubit relaxation/excitation experiments
involving non-equilibrium quasiparticles \cite{Martinis09}, will
attempt to relax the qubit energy splitting which is $\simeq\omega_{p}$
into quasiparticles, hence, only the main band is relevant. For a
quantitative estimate we can identify
\begin{equation}
\rho_c=\sqrt{\frac{2E_{c}}{E_{J}}}.
\end{equation}
where the charging energy is defined for electrons, i.e. $E_c=e^2/2C$. 

{Note the $\rho_c^2$ in the the second term of eq. (\ref{eq. gamma1}) provides the coefficient $2 E_c/E_J$ for the $\Gamma_{\rm dr,1}$, which makes this term equivalent eq. (2) in \cite{Martinis09}. 
 For a transmon, $E_{J}/E_{c}\approx30$ and for a flux qubit $\approx50$,
making this correction large enough to be visible in a reduction of
the quasiparticle rate. More explicitly, for a transmon we get a difference of
around 7\% from the perturbative approximation ($1-e^{-\rho_c/2} \approx \rho_c/2$)
and for a flux qubit we get a deviation of around 5\%. For a traditional
charge qubit, $\rho_c$ is large, leading to a different regime where
linearization of phase fluctuations is impossible. From
\cite{Charge03}, $E_{J}/E_{c}\approx0.35$ and with this caveat we can estimate a large deviation of around 40\% from the perturbative
approximation.

\subsection{Overdamped environmental mode \label{sec:overdamped}}

We are now studying the opposite case, an environmental mode that
is overdamped by an external impedance. Overdamping means the width
of the resonance in $Z_{{\rm eff}}$, eq. \ref{eq:lorentz_spectrum}
is larger than its frequency, i.e. $\omega_{p}RC\ll1$ or equivalently
$R\ll Z_{0}$. In that case, to lowest order in the small parameter
$r=\frac{R}{Z_{0}}$ is, the poles of $Z_{{\rm eff}}$, eq. \ref{eq:lorentz_spectrum}
are at $\omega=\pm i\gamma$ and at $\omega=\pm i\gamma r^{2}$ where
$\gamma=1/RC$. Note that unlike other work for overdamped oscillators
\cite{Weiss99}we keep the next-to leading order which keeps the second
set of poles to zero, rendering the quasiparticle current nonzero
in view of eq. (\ref{eq: ptot_final}). This will have an important
consequence later on. We can now, in the overdamped case, rewrite
eq. (\ref{eq:lorentz_spectrum}) as 
\begin{equation}
Z_{{\rm eff}}\simeq R\gamma^{2}\frac{1}{1-r^{4}}\left[\frac{1}{\omega^{2}+\gamma^{2}}-\frac{r^{4}}{\omega^{2}+\gamma^{2}r^{4}}\right].\label{eq:overdamped_impedance}
\end{equation}

This can be read as the difference between two Ohmic environments
with a Drude cutoff \cite{Weiss99}. This allows us to rely on known
results \cite{Ingold92,Schoen98,Weiss99,PRL01} for computing $J(t)$.
Here, we focus on the zero-temperature regime. We find for the zero-point
term $S(0)=-\frac{2R}{R_{K}}\log r$.

Now, interestingly, this will be a large positive term for $r\rightarrow0$
which can be achieved by going to small $\omega_{p}$, i.e. the $uv$
-term in eq. \ref{eq: ptot_final} will be suppressed by a factor
$r^{4R/R_{K}}$ to a very small value.

It is known \cite{Schoen98}is that for $Z_{d}=\frac{R_{0}}{1+\omega^{2}/\omega_{0}^{2}}$
we have 
\begin{equation}
\dot{J_{S}}(t)=\frac{R_{0}\omega_{0}}{R_{K}}\left[e^{-\omega_{0}\tau}E_{1}(-\omega_{0}\tau)-e^{\omega_{0}\tau}E_{1}\left(\omega_{0}\tau\right)\right]
\end{equation}
and that for long times, $\omega_{0}\tau\rightarrow\infty$ this reduces
to $J_{S}(t)=-\frac{2R_{0}}{R_{K}}\left(\log\omega_{c}t+\gamma_{e}+i\frac{\pi}{2}\right)$
where $\gamma_{e}\simeq0.5772$ is the Euler-Mascheroni constant.
For short times, $\omega_{0}\tau\ll1$, we find $J(t)=i\pi\alpha\omega_{0}\pi t$.
We can now find $P(E)$ in three regimes. For $E\ll\gamma,\gamma r^{2}$,
both terms in eq. \ref{eq:overdamped_impedance} should be treated
in the long-time limit. In lowest order in $r$ we find 
\begin{equation}
J(t)=-\frac{4R}{R_{K}}\log r
\end{equation}
i.e., no time-dependence, in agreement with the fact that the environment
is super-Ohmic at these low frequencies, see eq. \ref{eq:overdamped_impedance}.
This leads to $P_{s}(E)=r^{\rho_c}\delta(E)$.

In the intermediate regime, $\gamma r^{2}\ll E\ll\gamma$, we can
combine short and long-time limit as 
\begin{equation}
J(t)=-\frac{R_{0}}{R_{K}}\left[\log\gamma t+\gamma_{e}+i\frac{\pi}{2}-i\pi\gamma r^{2}t\right]
\end{equation}

This leads to an Ohmic $P(E)$ that is energetically shifted by $\delta E=\frac{R_{0}}{R_{K}}\pi\gamma r^{2}=\frac{R^{2}}{LR_{K}}$.
This results in 
\begin{eqnarray}
P(E) & = & \frac{e^{-2\gamma_{e}R_{K}/R}}{\Gamma(2R_{K}/R)}\frac{1}{E}\left(\frac{\pi R_{K}}{R}\frac{E}{E_{c}}\right)^{2R_{K}/R}\nonumber \\
 &  & ~~~~~~\times\left(1+\left(\frac{2R_{K}}{R}-1\right)\frac{\delta E}{E}\right)
\end{eqnarray}

Finally, at large energies hence entirely short times and large energies,
$E\gg\gamma$, we can approximate 
\begin{equation}
J(t)\simeq\frac{i\pi}{C}t
\end{equation}
hence leading to a simple capacitive contribution from the junction's
charging energy and 
$P(E,E^{\prime}) = e^{-\frac{S(0)}{4}} \left(\left(u^{2}+v^{2} \right)e^{\frac{S(t)}{4}} - 2uve^{-\frac{S(t)}{4}}
\right)   \delta\left(E^{\prime}-E-e^{2}/2C\right). $

\section{Conclusion}

In conclusion, we have developed the theory of quasiparticle tunneling
for superconducting tunnel junctions in an arbitrary linear dissipative
environment. We worked out the unperturbed tunneling rate of nonequilibrium
quasiparticles in the junction. In the perturbation regime P(E) governs
processes of an electron or hole tunneling, which are represented
by the diagonal transition matrix elements in charge space. We showed
that processes that create charge superpositions are additionally
supressed by zero-point fluctuations of the phase. For the case of
low damping, quasiparticle tunneling is exchanging energy with the
plasma mode in the form of a sequence of sidebands.

\section{Acknowlegement} 

This work was supported by NSERC through the discovery grants program, Office of the Director of National
Intelligence (ODNI), Intelligence Advanced Research Projects Activity
(IARPA). AS gratefully acknowledges support from Perimeter Institute for Theoretical Physics where
most of his contribution was worked out. Discussions with
J.M. Martinis, A.J. Leggett, and G. Catelani are gratefully acknowledged.

\section*{References}
\bibliography{frankslibrary,frankspapers,PaperLib,MohammadLibrary}
\bibliographystyle{prsty}

\end{document}